\documentclass[letterpaper, 10 pt, draftcls, onecolumn, conference]{ieeeconf}

\IEEEoverridecommandlockouts                              % This command is only needed if
\pdfoutput=1
\usepackage{amssymb}
\usepackage{subfigure}% Support for small, `sub' figures and tables. Your choice of alternative may be preferred instead.
\usepackage{graphicx}
\usepackage{enumerate}
\usepackage{amsmath}
\usepackage{amsfonts}
\usepackage{algorithm}
\usepackage{algpseudocode}
\usepackage{url}
\usepackage{epstopdf}
\usepackage{multirow}
\usepackage{authblk}
\usepackage{xcolor}
\usepackage{epstopdf}
\usepackage{tabularx}
\usepackage{stmaryrd}
\usepackage{setspace}
\usepackage{tcolorbox}
\usepackage{soul}% 高亮文字
\usepackage{colortbl}

\newtheorem{theorem}{Theorem}[section]
\newtheorem{example}[theorem]{Example}
\newtheorem{corollary}[theorem]{Corollary}
\newtheorem{proposition}[theorem]{Proposition}
\newtheorem{definition}[theorem]{Definition}
\newtheorem{lemma}[theorem]{Lemma}

\newcommand{\prop}{\begin{proposition}}
\newcommand{\eprop}{\end{proposition}}
\newcommand{\thm}{\begin{theorem}}
\newcommand{\ethm}{\end{theorem}}
\newcommand{\dfn}{\begin{definition}}
\newcommand{\edfn}{\end{definition}}
\newcommand{\exm}{\begin{example}}
\newcommand{\eexm}{\end{example}}
\newcommand{\coro}{\begin{corollary}}
\newcommand{\ecoro}{\end{corollary}}
\newcommand{\lem}{\begin{lemma}}
\newcommand{\elem}{\end{lemma}}
\newcommand{\prof}{\begin{proof}}
\newcommand{\eprof}{\end{proof}}
\newcommand{\red}{\color{red}}

\title{\LARGE \bf
Benchmarks of Extended Basis Reachability Graphs
}

\author{Yin Tong\\
{\normalsize(email:yintong@swjtu.edu.cn)}\\
\emph{\normalsize School of Information Science and Technology, Southwest Jiaotong University, China }
}

\begin{document}

\maketitle
\thispagestyle{empty}
\pagestyle{empty}

In this note, we want to provide a comparison among the efficiency of different approaches for the verification of $K$-step and infinite-step opacity based on three different graphs:
the Extended Basis Reachability Graph (EBRG) proposed in \cite{ref1}, the Basis Reachability Graph (BRG) when applicable \cite{ref1}, and the Reachability Graph (RG). All the algorithms are coded in Python\footnote{One can request the source file of the code by contacting the author through email.}, and the numerical simulations are run on a laptop with AMD Ryzen 9 3900X 3.79 GHz CPU and 16 GB RAM.

To make the comparison meaningful, according to the theory proposed in \cite{ref1}, the considered Petri net benchmarks should satisfy the following constraints:
\begin{enumerate}
  \item The Petri net system should be live and bounded.
  \item The Petri net should not be acyclic.
  \item There must be at least one unobservable transition.
  \item The unobservable subnet should be acyclic.
\end{enumerate}

\section{Benchmark 1}

\begin{figure}
  \centering
  \includegraphics[width=0.5\textwidth]{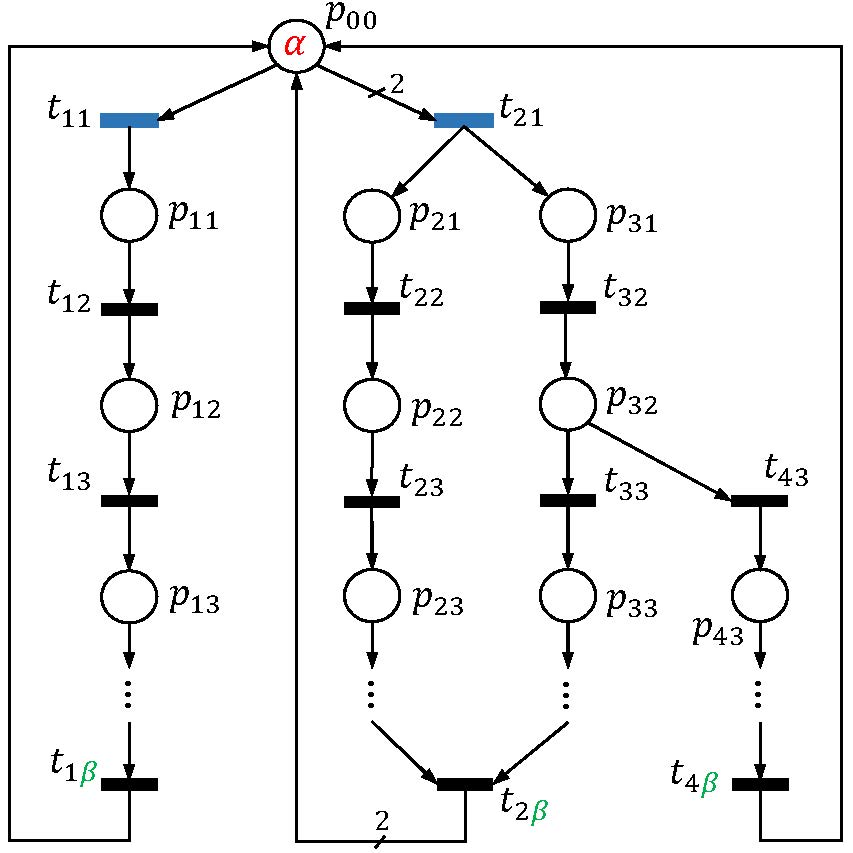}\\
  \caption{Benchmark 1.}\label{fig:ben1}
\end{figure}

The first benchmark is the Petri net with parameters $\alpha$ and $\beta$ in Fig.~\ref{fig:ben1}. The Petri net contains 4 elementary cycles:
\begin{itemize}
  \item Cycle 1: $p_{00}$ - $t_{11}$ - $p_{11}$ - $\ldots$ - $p_{1\beta-1}$ - $t_{1\beta}$ - $p_{00}$
  \item Cycle 2: $p_{00}$ - $t_{21}$ - $p_{21}$ - $\ldots$ - $p_{2\beta-1}$ - $t_{2\beta}$ - $p_{00}$
  \item Cycle 3: $p_{00}$ - $t_{21}$ - $p_{31}$ - $\ldots$ - $p_{3\beta-1}$ - $t_{2\beta}$ - $p_{00}$
  \item Cycle 4: $p_{00}$ - $t_{21}$ - $p_{31}$ - $t_{32}$ - $p_{32}$ - $t_{43}$ - $p_{43}$ - $\ldots$ - $p_{4\beta-1}$ - $t_{4\beta}$ - $p_{00}$
\end{itemize}

In each elementary cycle, there are $\beta$ transitions and $\beta$ places. In total, there are $4\beta-4$ transitions and $4\beta-5$ places. The parameter $\alpha$ denotes the number of tokens in place $p_{00}$, i.e., the initial marking of the net is $M_0=\alpha p_{00}$.

Then, for different values of $\beta$ and $\alpha$, the Petri net structure and the initial marking change. Therefore, Fig.~\ref{fig:ben1} illustrates \emph{a class of Petri net systems}, instead of one Petri net system. Since most of the Petri net systems satisfying the above four constraints are composed by several elementary cycles, the class of nets in Fig. 1 is quite general.

To satisfy Constraints 3) and 4), transitions $t_{11}$ and $t_{21}$ are observable transitions while all other transitions are unobservable. Later on, we will add (randomly) more observable transitions to test the change of the size of RG, BRG and EBRG. We denote $\lambda$ the number of observable transitions in the net system.

\subsection{Test Cases}

The main advantage of using BRG and EBRG over RG to verify $K$-step and infinite-step opacity is that there is no need to compute the whole state space but only a subset of the reachable markings (which are called basis markings in the BRG case, and extended basis markings in the EBRG case, respectively). When the computation of all the reachable markings is not possible in practice, the BRG and EBRG-based approaches may provide a solution, especially for large-scaled Petri nets.

The test cases should show how the size of the Petri net systems affects the number of reachable, basis reachable, and extended basis reachable markings. We are also interested in how the number of reachable, basis reachable and extended basis reachable markings changes with an increasing number of observable transitions. Therefore, for the class of nets in Fig.~\ref{fig:ben1}, we consider three sets of test cases summarized in Fig.~\ref{fig:ben1}.

\begin{figure}
  \centering
  \includegraphics[width=0.6\textwidth]{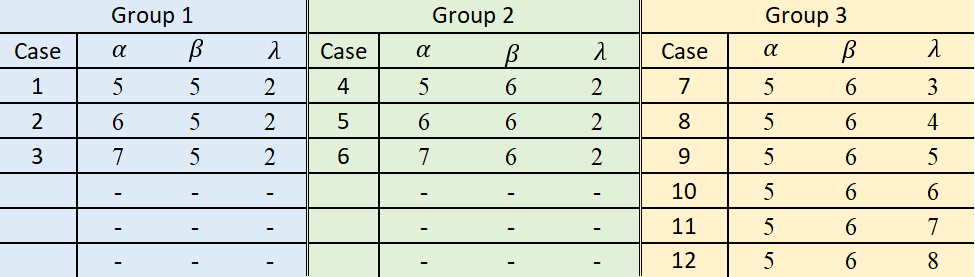}\\
  \caption{Three groups of test cases for Benchmark 1.}\label{fig:test_cases1}
\end{figure}

In Group 1, given $\beta=5$ and $\lambda=2$, the net structure does not change but the number of tokens initially contained in $p_{00}$ increases. Group 2 differs from Group 1 for the addition of 4 unobservable transitions and 4 places with $\beta=6$. Finally Group 3 differs from Group 2 for the presence of more observable transitions, which are randomly chosen. In total, we consider 12 test cases.

The size of the EBRG may also depend on the set $S\subseteq R(N,M_0)$ of secret markings. However, with different values of $\alpha$ and $\beta$, the set of reachable markings are different. Therefore, it is impossible to consider the same set $S$ of secret markings for all the cases. To be fair, the secret is taken as the set of markings reachable by firing two given sequences of transitions, namely we assume $$S=\{M_0,M_1,M_2\},$$ where $M_0$ is the initial marking, $M_0 [t_{21} t_{21} \rangle M_1 $, and $M_0 [t_{21} t_{22} t_{32} t_{33} \rangle M_2$. Note that one basis marking in the secret does not satisfy Assumption~A3 in \cite{ref1}. Otherwise, there is no need to construct the EBRG. Since $\alpha$ takes 3 different values in the 12 test cases, there are 3 sets of secret markings corresponding to $\alpha$ equal to 5, 6, and 7, respectively:
\begin{itemize}
  \item $S_1=\{M_0=5p_{00},M_1=3p_{00}+p_{21}+p_{31},M_2=3p_{00}+p_{22}+p_{33}\}$,
  \item $S_2=\{M_0=6p_{00},M_1=4p_{00}+p_{21}+p_{31},M_2=4p_{00}+p_{22}+p_{33}\}$,
  \item $S_3=\{M_0=7p_{00},M_1=5p_{00}+p_{21}+p_{31},M_2=5p_{00}+p_{22}+p_{33}\}$.
\end{itemize}

\subsection{Numerical Results}
\begin{figure}
  \centering
  \includegraphics[width=0.75\textwidth]{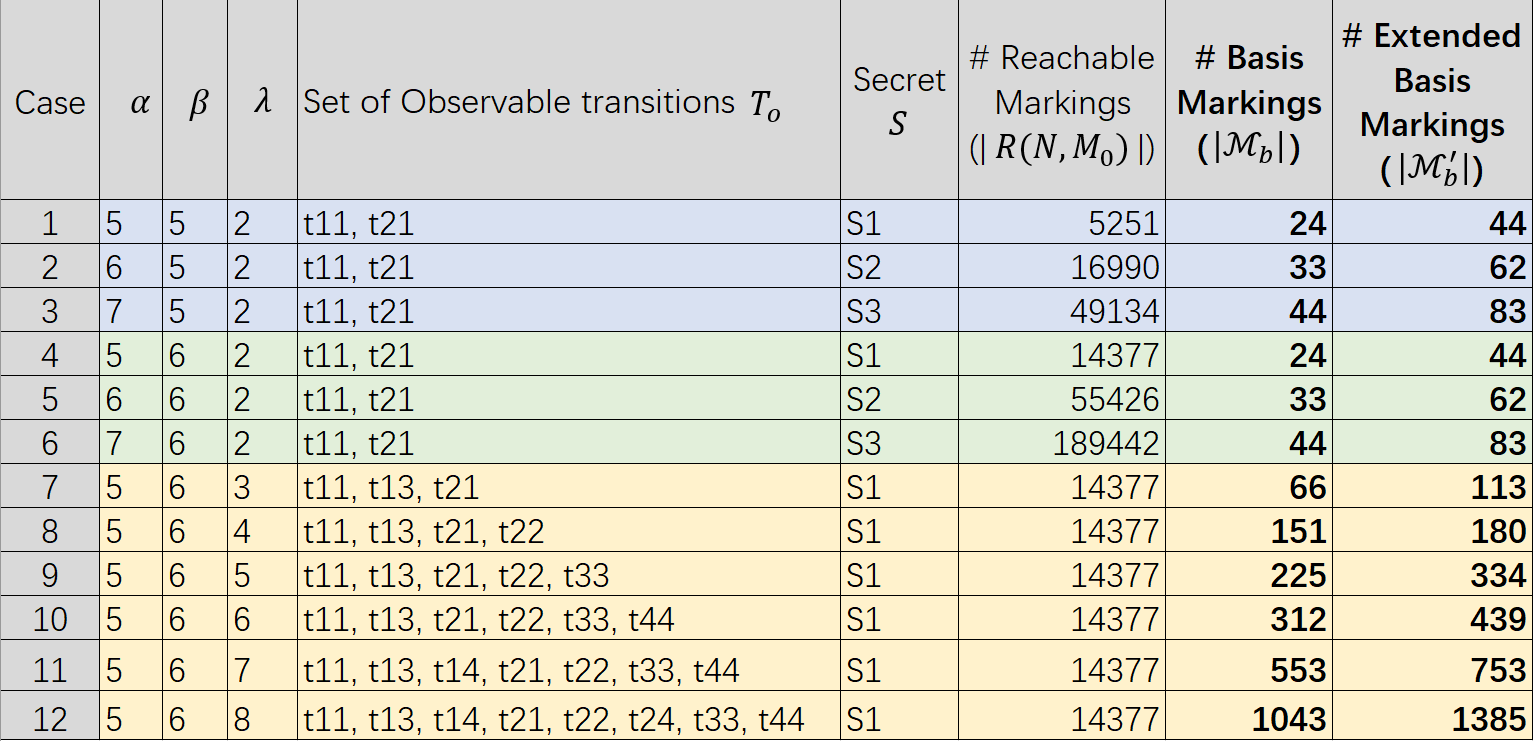}\\
  \caption{The number of reachable, basis reachable, and extended basis reachable markings of Benchmark 1.}\label{fig:no_marking1}
\end{figure}

\begin{figure}
  \centering
  \includegraphics[width=0.6\textwidth]{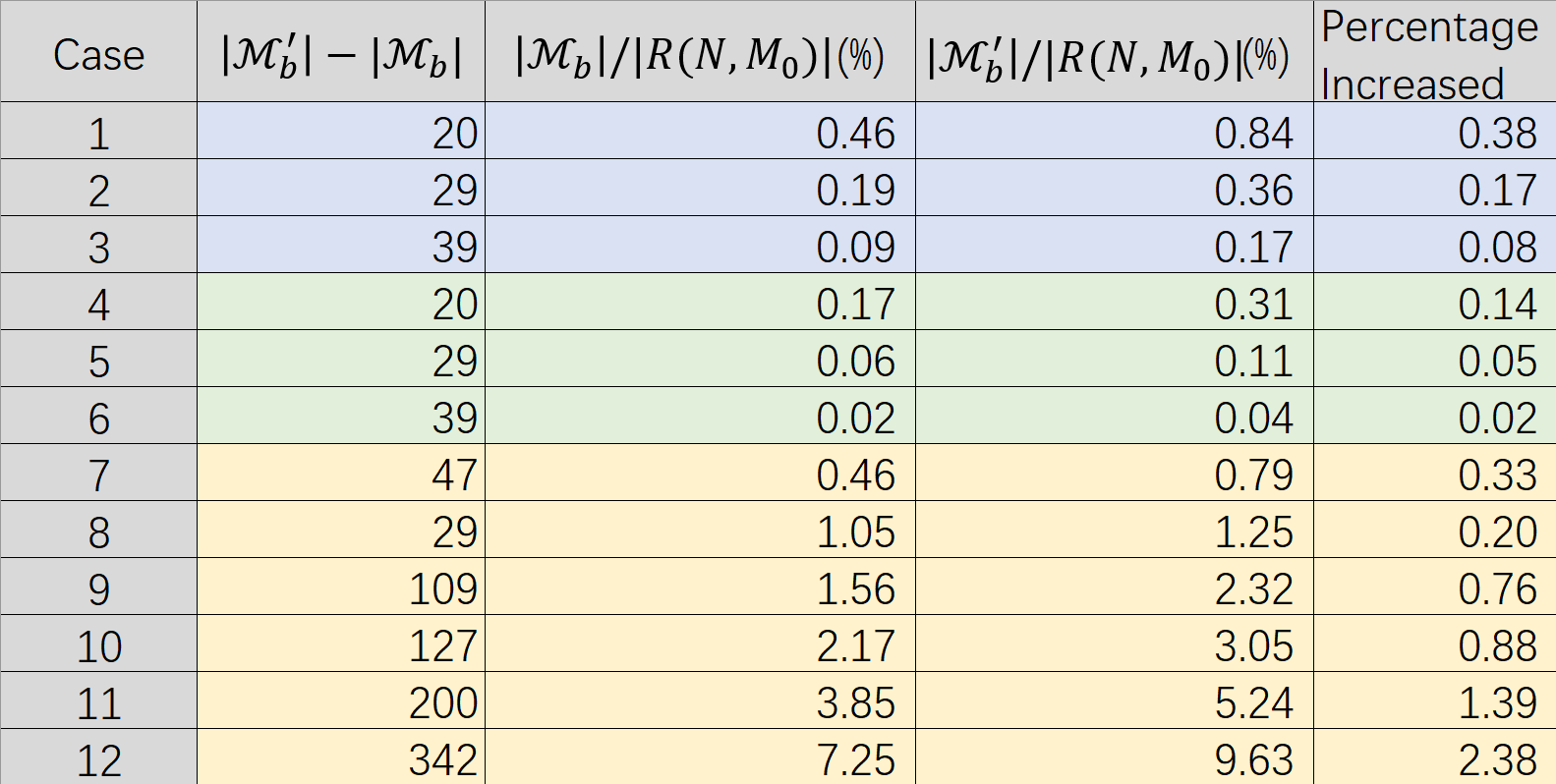}\\
  \caption{Comparison of number of markings in RG, BRG, and EBRG of Benchmark 1.}\label{fig:comp1}
\end{figure}

\begin{figure}[h]
  \centering
  \includegraphics[width=0.45\textwidth]{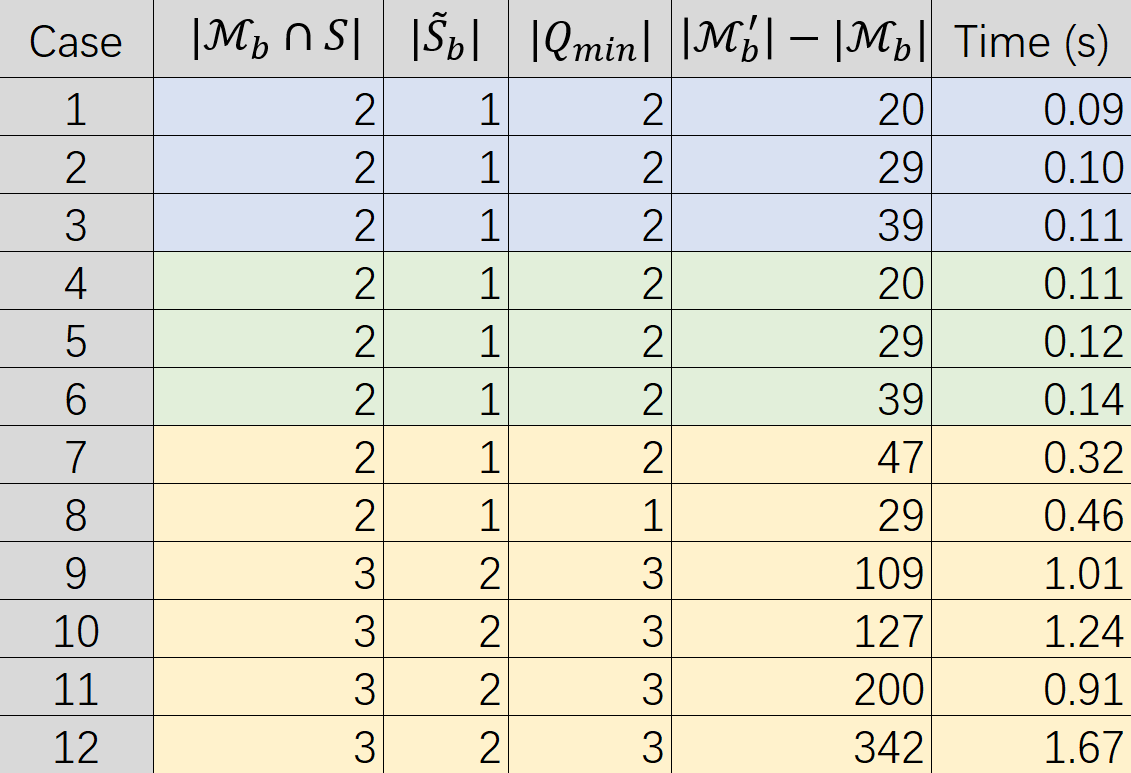}\\
  \caption{Time cost and calculation of intermediate sets of markings for constructing the EBRG of Benchmark 1.}\label{fig:stats_EBRG1}
\end{figure}

The number of reachable, basis, and extended basis markings for the 12 scenarios defined in the previous subsection are summarized in Fig.~\ref{fig:no_marking1}. In Fig.~\ref{fig:comp1}, a clear comparison between such numbers is proposed. Fig.~\ref{fig:stats_EBRG1} shows the time cost of computing the EBRG and the cardinality of sets ${\tilde S}_b$, $Q_{min}$, etc., whose calculation is necessary for constructing the EBRG.

We conclude that:
\begin{itemize}
  \item From Figs.~\ref{fig:no_marking1} and \ref{fig:comp1}: both BRG and EBRG are much smaller than the RG.
  \item From Figs.~\ref{fig:no_marking1} and \ref{fig:comp1}: with the increase of the size of the net system, the number of reachable markings increases much faster than that of basis and extended basis reachable markings.
  \item Comparing Group 3 with Case 4 in Fig.~\ref{fig:no_marking1}: the larger the number of observable transitions, the larger the number of basis markings and extended basis markings, and their increase speeds are still acceptable.
  \item From Fig.~\ref{fig:comp1}: the number of extended basis markings does not increase too much because of one basis marking not satisfying Assumption~A3.
  \item From Fig.~\ref{fig:stats_EBRG1}: the larger the number of basis markings not satisfying Assumption A3, the larger the number of  extended basis markings.
  \item From Fig.~\ref{fig:stats_EBRG1}: the EBRG can be constructed within 2 seconds.
\end{itemize}

\section{Benchmark 2}

\begin{figure}
  \centering
  \includegraphics[width=0.5\textwidth]{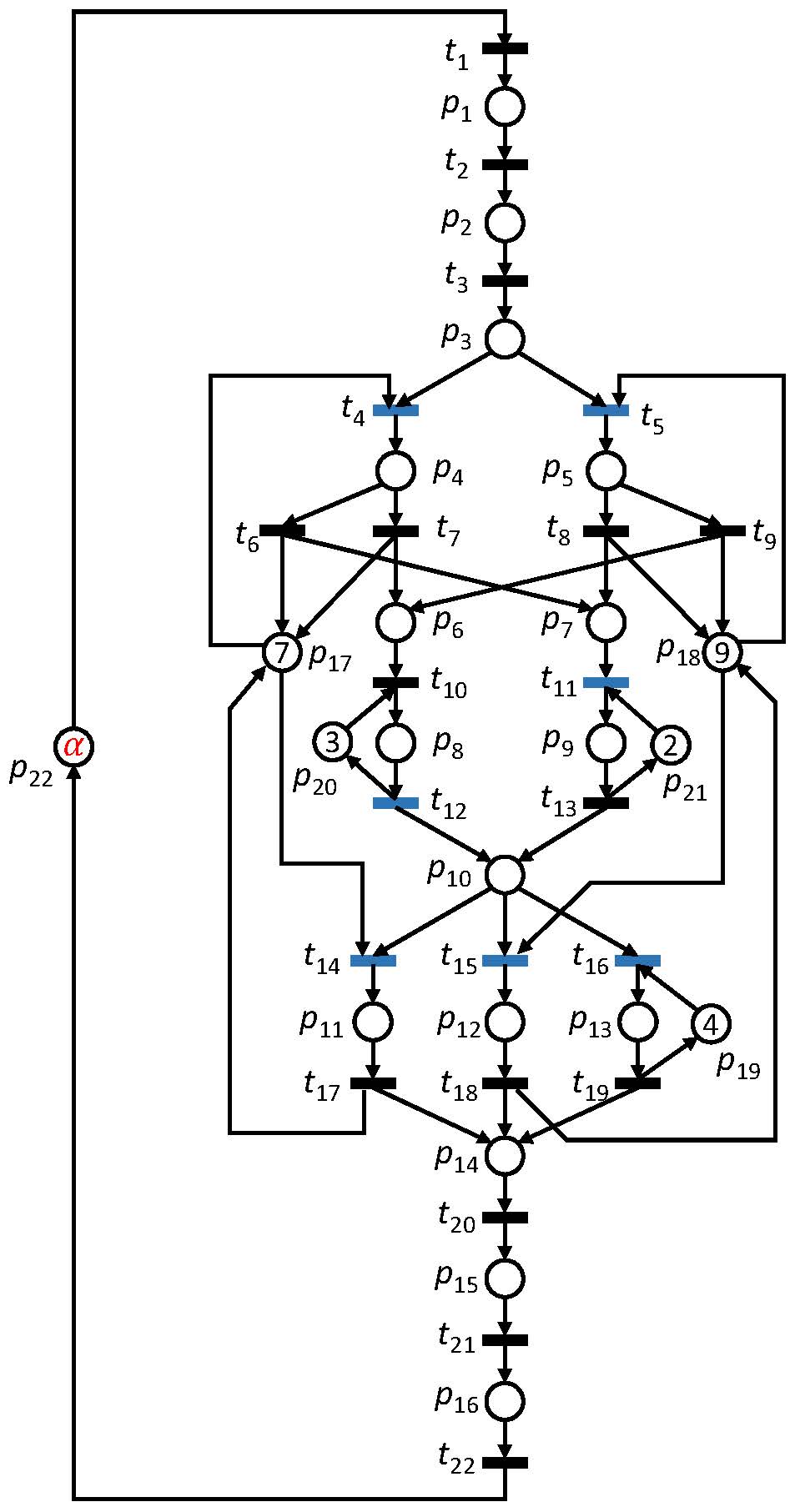}\\
  \caption{Benchmark 2.}\label{fig:ben2}
\end{figure}

The second benchmark is taken from \cite{ref2,ref3}, which models a Hospital Emergency Service System. The Petri net model is shown in Fig.~\ref{fig:ben2}. The meanings of places and transitions are described in \cite{ref3}. Place $p_{22}$ represents the maximum number of patients that can be accepted simultaneously in the Emergency Department (ED). Places $p_{17}$, $p_{18}$, $p_{19}$, $p_{20}$, $p_{21}$ are used to denote available ED medical doctors, ED registered nurses, transports, CT scanners, and X-ray, respectively. The initial marking of this net is
$$M_0=\alpha p_{22}+7p_{17}+9p_{18}+4p_{19}+3p_{20}+2p_{21},$$
where $\alpha$ is an integer number that denotes the capacity of the ED.

\subsection{Test Cases}
\begin{figure}
  \centering
  \includegraphics[width=0.8\textwidth]{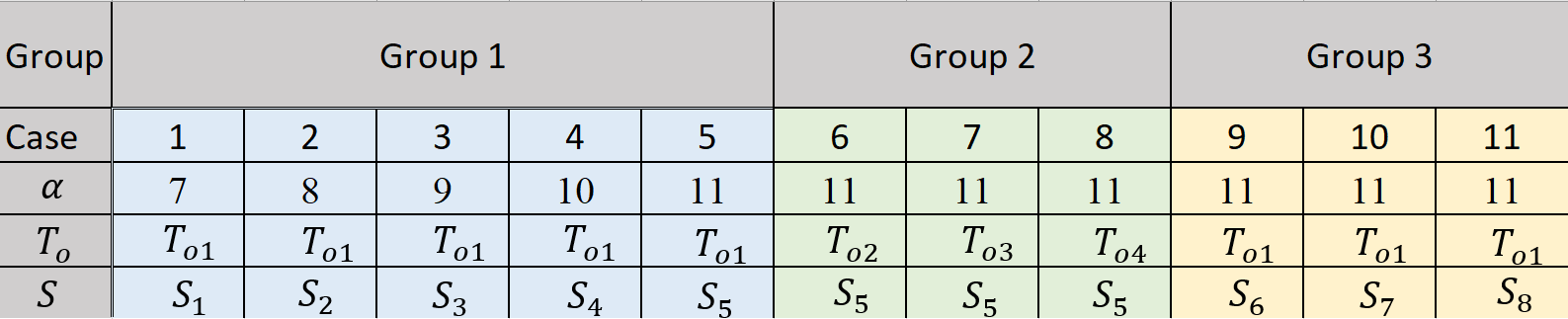}\\
  \caption{Three groups of test cases for Benchmark 2.}\label{fig:test_cases2}
\end{figure}

We define three groups of test cases that are summarized in Fig.~\ref{fig:test_cases2}. In total, we consider 11 test cases. The values of the sets $T_o$ of observable transitions and the sets $S$ of secret markings for different test cases are listed in Tables~\ref{tab:To} and \ref{tab:S}, respectively. The observable transitions are randomly chosen. Note that {\red sets} $S_1$ to $S_5$ have 4 secret markings, while sets $S_6$, $S_7$, and $S_8$ have 5, 6, and 7 secret markings, respectively. Note that since {\red the number of} basis markings is much smaller than that of reachable markings, the secret $S$ cannot be randomly chosen from $R(N,M_0)$, otherwise most likely there would be no need to construct the EBRG. Therefore, some of the secret markings are randomly chosen from the set of basis markings.

\begin{table}
\caption{Sets of observable transitions in Fig.~\ref{fig:test_cases2}.}
\centering\label{tab:To}
\begin{tabular}{c|l}
  \hline
  $T_o$  & Transitions\\
  \hline
  $T_{o1}$ & $\{t_4,t_5,t_{11},t_{12},t_{14},t_{15},t_{16}\}$ \\
  $T_{o2}$ & $\{t_1,t_4,t_5,t_{11},t_{12},t_{14},t_{15},t_{16}\}$\\
  $T_{o3}$ & $\{t_1,t_4,t_5,t_9,t_{11},t_{12},t_{14},t_{15},t_{16}\}$\\
  $T_{o4}$ & $\{t_1,t_4,t_5,t_9,t_{11},t_{12},t_{14},t_{15},t_{16},t_{21}\}$\\
  \hline
\end{tabular}
\end{table}

\begin{table}
\caption{Sets of secret markings in Fig.~\ref{fig:test_cases2}.}
\centering\label{tab:S}
\begin{tabular}{c|ll}
  \hline
  $S$  & Markings\\
  \hline
  $S_1$ & $\{M_0=7p_{22}$,\\
        & $M_1=6p_{22}+p_4+6p_{17}+9_p{18}+4p_{19}+3p_{20}+2p_{21}$, \\
        & $M_2=6p_{22}+p_6+7p_{17}+9p_{18}+4p_{19}+3p_{20}+2p_{21}$, \\
        & $M_3=5p_{22}+p1+p9+7p_{17}+9p_{18}+4p_{19}+3p_{20}+p_{21}$\}\\
  \hline
  $S_2$ & $\{M_0=8p_{22}$,\\
        & $M_1=7p_{22}+p_4+6p_{17}+9_p{18}+4p_{19}+3p_{20}+2p_{21}$, \\
        & $M_2=7p_{22}+p_6+7p_{17}+9p_{18}+4p_{19}+3p_{20}+2p_{21}$,\\
        & $M_3=6p_{22}+p1+p9+7p_{17}+9p_{18}+4p_{19}+3p_{20}+p_{21}\}$\\
  \hline
  $S_3$ & $\{M_0=9p_{22}$, \\
        & $M_1=8p_{22}+p_4+6p_{17}+9_p{18}+4p_{19}+3p_{20}+2p_{21}$,\\
        & $M_2=8p_{22}+p_6+7p_{17}+9p_{18}+4p_{19}+3p_{20}+2p_{21}$,\\
        & $M_3=7p_{22}+p1+p9+7p_{17}+9p_{18}+4p_{19}+3p_{20}+p_{21}\}$\\
  \hline
  $S_4$ & $\{M_0=10p_{22}$, \\
        & $M_1=9p_{22}+p_4+6p_{17}+9_p{18}+4p_{19}+3p_{20}+2p_{21}$, \\
        & $M_2=9p_{22}+p_6+7p_{17}+9p_{18}+4p_{19}+3p_{20}+2p_{21}$, \\
        & $M_3=8p_{22}+p1+p9+7p_{17}+9p_{18}+4p_{19}+3p_{20}+p_{21}\}$\\
  \hline
  $S_5$ & $\{M_0=11p_{22}$, \\
        & $M_1=10p_{22}+p_4+6p_{17}+9_p{18}+4p_{19}+3p_{20}+2p_{21}$, \\
        & $M_2=10p_{22}+p_6+7p_{17}+9p_{18}+4p_{19}+3p_{20}+2p_{21}$,\\
        & $M_3=9p_{22}+p1+p9+7p_{17}+9p_{18}+4p_{19}+3p_{20}+p_{21}\}$\\
  \hline
  \hline
  $S_6$ & $\{M_0$, \\
        & $M_1=10p_{22}+p_4+6p_{17}+9_p{18}+4p_{19}+3p_{20}+2p_{21}$,\\
        & $M_2=8p_{22}+p_9+p_{11}+p_{13}+6p_{17}+9p_{18}+3p_{19}+3p_{20}+p_{21}$, \\
        & $M_3=6p_{22}+p_5+p_9+p_{10}+2p_{12}+7p_{17}+6p_{18}+4p_{19}+3p_{20}+p_{21}$, \\
        & $M_4=4p_{22}+p_4+p_5+p_9+p_{10}+p_{11}+p_{12}+p_{13}+5p_{17}+7p_{18}+3p_{19}+3p_{20}+p_{21}\}$\\
  \hline
  $S_7$ & $\{M_0$, $M_1$, $M_2$, $M_3$, $M_4,$\\
        & $M_5=3p_6+2p_9+6p_{12}+7p_{17}+3p_{18}+4p_{19}+3p_{20}\}$\\
  \hline
  $S_8$ & $\{M_0$, $M_1$, $M_2$, $M_3$, $M_4$, $M_5$,\\
        & $M_6=7p_4+3p_5+p_{13}+6p_{18}+3p_{19}+3p_{20}+2p_{21}\}$\\
  \hline
\end{tabular}
\end{table}

For this benchmark, we want to test the performance of BRG and EBRG-based methods when RG cannot be constructed in practice (Group~1). In addition, we further test the increase of the number of basis and extended basis reachable markings with respect to the increase of the number of observable transitions (Group~2) and the increase of the number of secret markings (Group~3).

\begin{figure}
  \centering
  \includegraphics[width=0.55\textwidth]{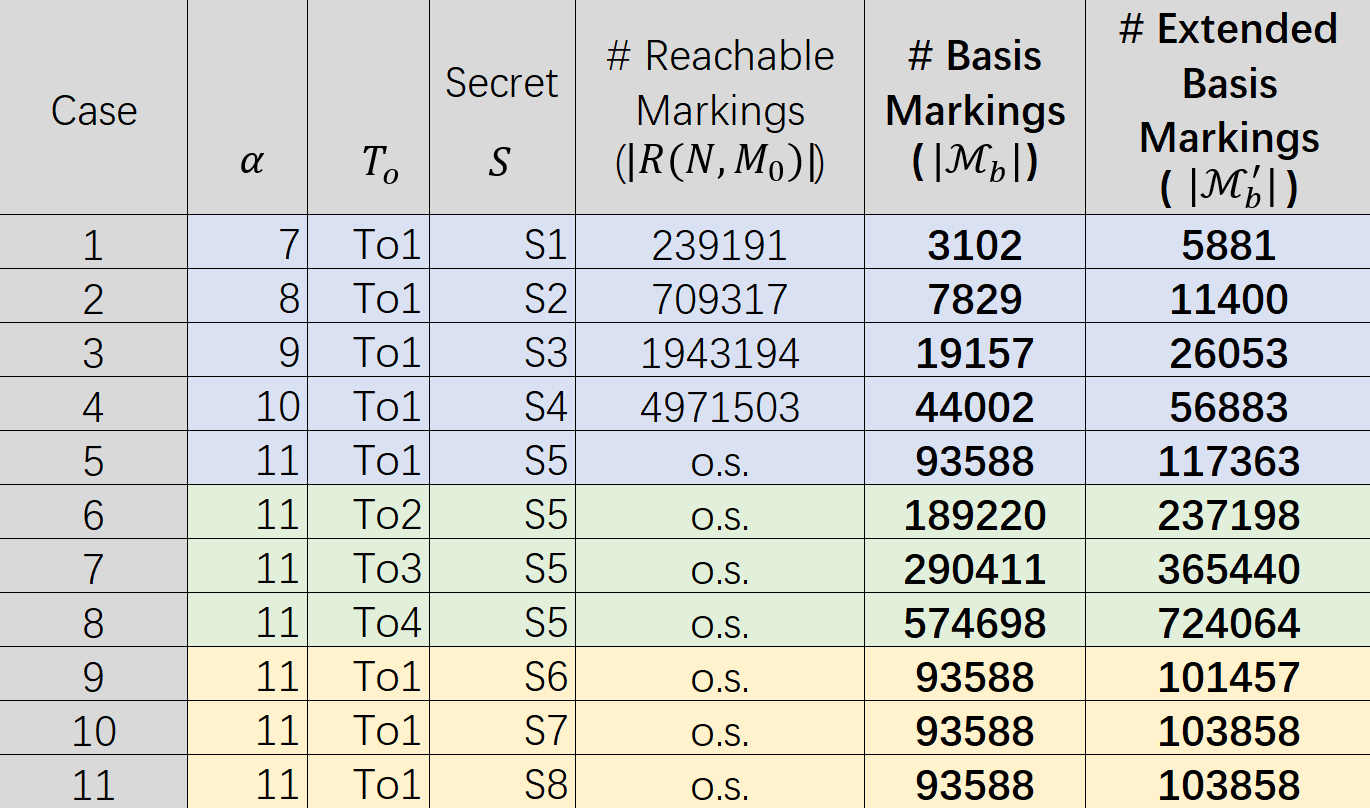}\\
  \caption{The number of reachable, basis reachable, and extended basis reachable markings of Benchmark 2, where ``o.s.'' means ``out of space''.}\label{fig:no_marking2}
\end{figure}

\begin{figure}[h]
  \centering
  \includegraphics[width=0.6\textwidth]{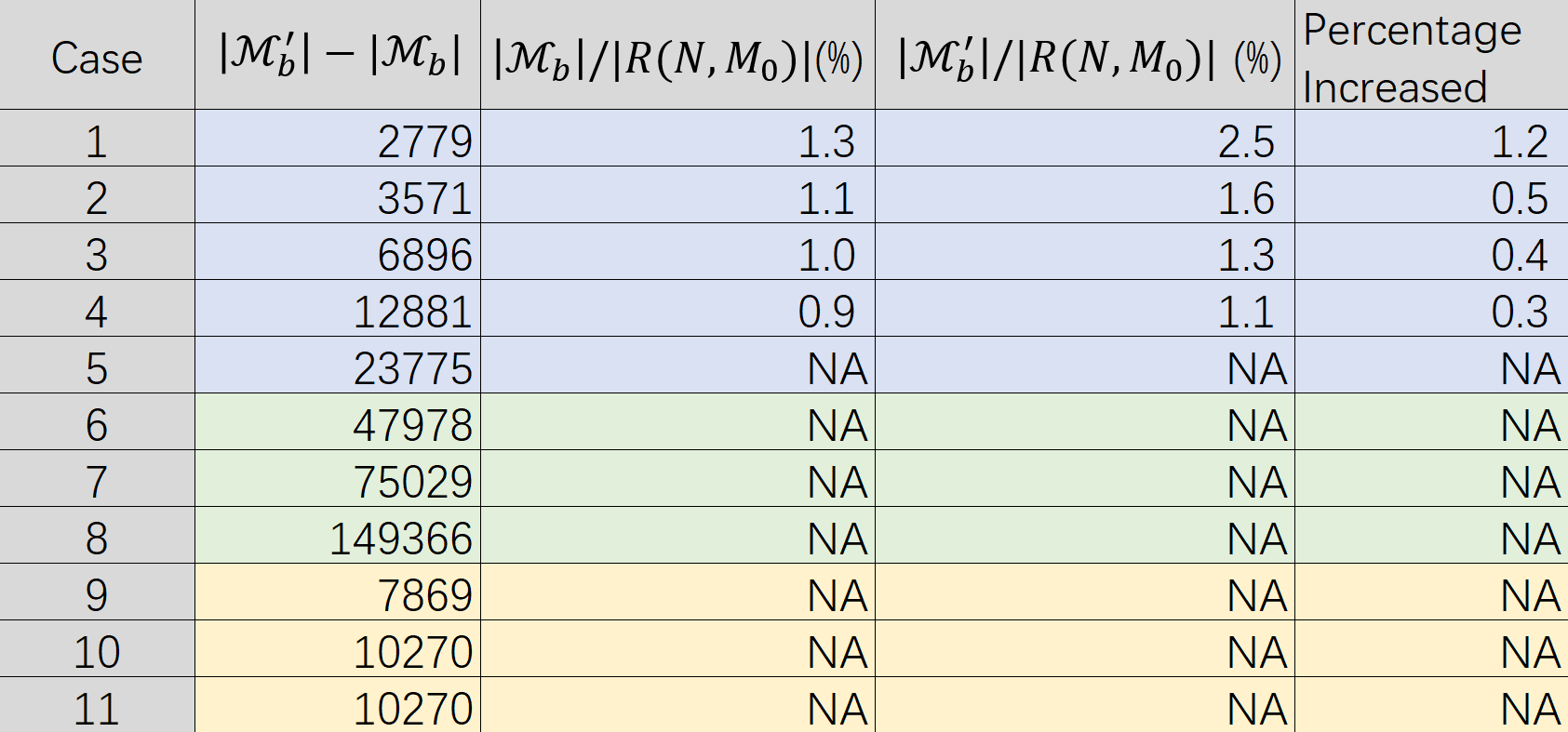}\\
  \caption{Comparison of number of markings in RG, BRG, and EBRG of Benchmark 2.}\label{fig:comp2}
\end{figure}

\begin{figure}[h]
  \centering
  \includegraphics[width=0.5\textwidth]{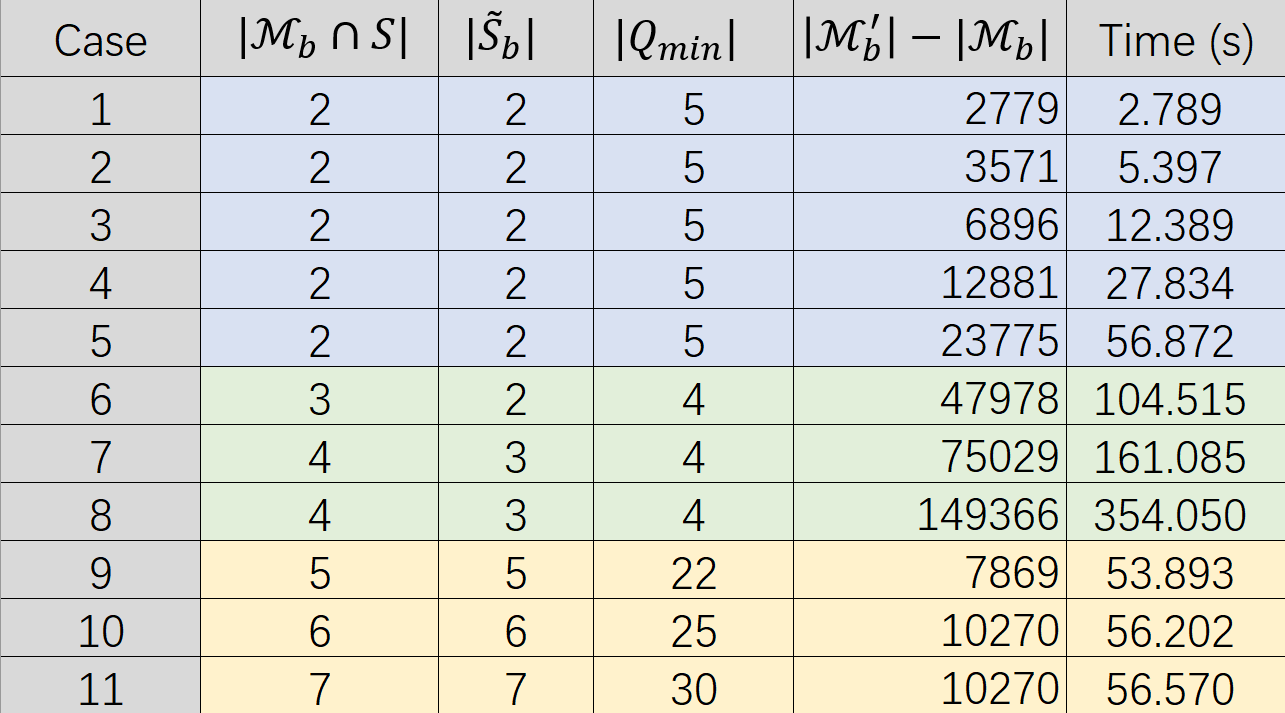}\\
  \caption{Time cost and calculation of intermediate sets of markings of constructing the EBRG of Benchmark 2.}\label{fig:stats_EBRG2}
\end{figure}

\subsection{Numerical Results}
The simulation results on Benchmark 2 are presented in Figs.~\ref{fig:no_marking2}, \ref{fig:comp2}, \ref{fig:stats_EBRG2}. From Fig.~\ref{fig:no_marking2} we can see that when $\alpha=11$, the program cannot obtain the reachable markings. However, the BRG and the EBRG can be computed and they are even much smaller than the RG in Case~2. Comparing the results in Group~2 and Group~3, we can see that the number of basis and extended basis markings increases faster because of the increase of the number of observable transitions rather than because of the increase of the number of secret markings, and the scale of increase is stable: between 1.5 and 2.0, which is quite slow.

From Fig.~\ref{fig:comp2}, we argue that the BRG and EBRG are both much smaller than the RG. From Fig.~\ref{fig:stats_EBRG2}, comparing the results in Group~3 with the results of Case~5 in Group~1, we found that the increase of the number of secret markings does not necessarily make the EBRG larger. It is also shown that the construction of the EBRG can be finished very quickly, always within 1 minute.

\end{document}